\documentclass[12pt]{iopart}
\usepackage{iopams}

\usepackage{graphicx}

\newcommand{\ket}[1]{\vert#1 \rangle}
\newcommand{\bra}[1]{\langle#1 \vert}
\newcommand{\expect}[1]{\langle #1 \rangle}

\begin{document}
\title{Protecting subspaces by acting on the outside}

\author{Jonathan Busch and Almut Beige}
\address{The School of Physics and Astronomy, University of Leeds, Leeds, LS2 9JT, United Kingdom}
\ead{a.beige@leeds.ac.uk}

\begin{abstract}
Many quantum control tasks aim at manipulating the state of a quantum mechanical system within a finite subspace of states. However, couplings to the outside are often inevitable. Here we discuss strategies which keep the system in the controlled subspace by applying strong interactions onto the outside. This is done by drawing analogies to simple toy models and to the quantum Zeno effect. Special attention is paid to the constructive use of dissipation in the protection of subspaces. 
\end{abstract}

\section{Introduction}

Quantum control techniques like Hamiltonian engineering are very successful when manipulating finite-dimensional Hilbert spaces \cite{control,control22,control33}. However, quantum control tasks often require the control of an infinite-dimensional state space. This applies for example, when the quantum system of interest couples to an infinitely large reservoir or when it contains bosonic modes with infinitely many states. In such a situation one should take advantage of mechanisms which restrict the time evolution of the system effectively onto a finite-dimensional space. The corresponding evolution can then be used to engineer appropriate control sequences \cite{Rabitz,pulses}.  

\begin{figure}[b]
\begin{center}
\resizebox{20pc}{!}{\rotatebox{-90}{\includegraphics{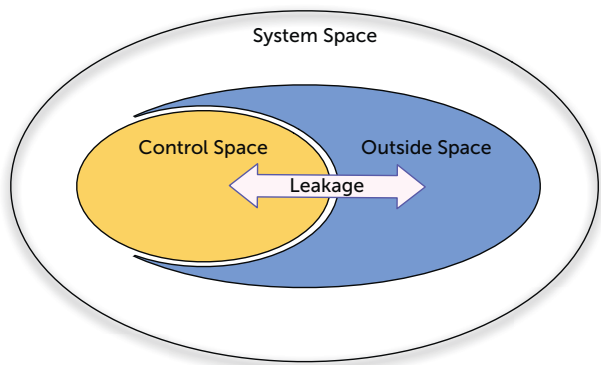}}}
\end{center}
\caption{Illustration of the control problem. We wish to control the system evolution within a subspace (yellow) of the total space (white) whilst there is coupling to an external subspace (blue).} \label{tony_0}
\end{figure}

The protection of finite-dimensional subspaces against the leakage of population (cf.~Fig.~\ref{tony_0}) can be achieved in many different ways. One approach is to use Hamiltonians which act only on small subsets of states and naturally restrict the time evolution of the system onto a finite-dimensional state space. This technique is used for example in ion trap quantum computing, where gate operations are realised by applying the required interactions in successive steps \cite{chitra}. During each time step, the number of phonons in the system increases at most by one. The populated state space remains finite and the excitation of coherent phonon states has been avoided. At the end of every control sequence, the ions return into a state with zero phonon excitation \cite{blatt,wineland}. 

Another very efficient tool for the protection of subspaces against leakage errors are stimulated Raman adiabatic passages (STIRAP) \cite{STIRAP,STIRAP2}. These employ the adiabatic theorem to induce transitions between states with no direct coupling between them. In composite quantum systems, like atoms which move slowly through an optical cavity, STIRAP can create ground state entanglement without populating excited electronic states and without creating photons inside the resonator \cite{carsten}. An alternative approach for protecting subspaces against leakage errors is to use especially designed impulsive pulse sequences. This technique has initially been developed to minimise radiation damage when exciting specific ground-state vibrational modes of molecules \cite{LCT}. A generalisation of both strategies is to simply use numerical simulations which impose state dependent constraints to design optimal control sequences \cite{Koch}.

In this paper we discuss a very different strategy for protecting the controlled subspace against leakage. Instead of imposing a well designed dynamics onto its states, we consider strong interactions which act {\em only} on the outside. If these interactions introduce a time scale into the system which is much shorter than the time scale on which the leakage of population out of the controlled subspace would occur in the unprotected case, one can show that these unwanted transitions become strongly inhibited \cite{Shore,Peres,Facchi2}. This approach has many similarities with bang-bang and its generalisation dynamical decoupling \cite{viola,viola2,zanardi} which also interrupt a relatively slow evolution with strong interactions. 

The second half of the paper pays special attention to the constructive use of dissipation in the protection of subspaces against population leakage and assumes non-zero spontaneous decay rates of the outside states. We are interested in cases, where the time evolution of the system can be understood in terms of rapidly-repeated measurements whether the system remains in the controlled subspace or not \cite{carsten,Zurek,Dugic,Beige00,Pachos,ions,Pechen}. These force the system to remain there much longer than in the unprotected case which can be understood in terms of the quantum Zeno effect \cite{misra,tony,zeno}. 
Notice that this approach provides a build-in error detection mechanism, when it is possible to register and act upon unwanted measurement outcomes. Moreover, as we shall see below, dissipation can protect subspaces against leakage errors even in situations, where other methods would simply fail. 

Refs.~\cite{Facchi4,Facchi3} discuss similarities between bang-bang, dynamical decoupling, the protection of subspaces with strong interactions, and the protection of subspaces using dissipation. The authors conclude that all these approaches are essentially equivalent, since all of them can be understood in terms of the quantum Zeno effect \cite{misra}. The purpose of the present manuscript is to give more insight into the underlying processes. This is done by analysing relatively simple toy models which allow us to compare the above mentioned methods qualitatively as well as quantitatively. 

This paper is organized in five sections. In the next section, we consider a two-level system with resonant coherent coupling to obtain information about the expected leakage rates in a simple unprotected subspace scenario. In Section \ref{pro1}, we extend the outside space and show how strong interactions acting on the outside space can be used (or how they should not be used) to protect the controlled subspace against leakage error. Section \ref{dissi} analyses closely related level schemes but with non-zero spontaneous decay rates. An example is given where dissipation results in the protection of a subspace which would not be there otherwise. We finally summarise our findings in Section \ref{conc}.

\section{An unprotected subspace} \label{unpro}

\begin{figure}[t]
\begin{center}
\resizebox{13pc}{!}{\rotatebox{-90}{\includegraphics{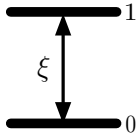}}}
\end{center}
\caption{Toy model illustrating the leakage of population from an unprotected controlled subspace (represented by $|0 \rangle$) with coupling strength $\xi$ into an outside space (represented by $|1 \rangle$).} \label{tony_1}
\end{figure}

Let us first consider a case where no effort is made to protect a controlled subspace from leaking population into outside states. For simplicity, we assume that the controlled subspace contains only a single state $|0 \rangle$. As shown in Fig.~\ref{tony_1}, there is moreover only one relevant state outside the controlled subspace which we denote by $|1 \rangle$. The leakage of population from level 0 into level 1 could be due to resonant interactions (like a laser field). Although this is an almost trivial case, the analysis of the time evolution of this level scheme introduces the relevant time scales of the system. This will enable us later to characterise and to compare the effectiveness of different strategies for the protection of controlled subspaces against leakage errors.

In the following, we assume that the laser is in resonance with the 0--1 transition and denote its (real) Rabi frequency by $\xi$. Moreover, $\hbar \omega_i$ denotes the energy of states $|i \rangle$. Then the system Hamiltonian in the usual rotating wave and dipole approximation can be written as
\begin{eqnarray} \label{H}
H &=& \hbar \xi \, {\rm e}^{{\rm i} \omega_\xi t} \, |0 \rangle \langle 1| + {\rm h.c.} + \sum_{i=0}^1 \hbar \omega_i \, |i \rangle \langle i| 
\end{eqnarray}
with $\omega_\xi \equiv \omega_1 - \omega_0$. To solve the corresponding time evolution, we first change into an the interaction picture with respect to $H_0 = \sum_{i=0}^1 \hbar \omega_i \, |i \rangle \langle i|$. This transfers the Hamiltonian (\ref{H}) into the interaction Hamiltonian
\begin{eqnarray}\label{HI_1}
H_{\rm I} &=& \hbar \xi \, |0 \rangle \langle 1| + {\rm h.c.}
\end{eqnarray}
To estimate the leakage rate of the controlled subspace in this case, we now calculate the population $P_0$ in $|0 \rangle$ at time $t$, given that the system was initially in $|0 \rangle$. 

One way of doing this is to consider the usual Pauli operators $\sigma_2$ and $\sigma_3$, 
\begin{eqnarray} \label{defq}
\sigma_2 = -{\rm i} \, ( \, |0 \rangle \langle 1| - |1 \rangle \langle 0| \, ) ~~ {\rm and} ~~
\sigma_3 = |0 \rangle \langle 0| - |1 \rangle \langle 1| \, ,
\end{eqnarray}
and to use the relation 
\begin{eqnarray}\label{dotA}
\langle \dot A \rangle &=& - {{\rm i} \over \hbar} \, \left \langle \left[ A , H_{\rm I} \right] \right \rangle 
\end{eqnarray}
for the time evolution of the expectation value of an operator $A$ in the interaction picture to obtain a closed set of rate equations. This yields the differential equations
\begin{eqnarray} \label{rateq}
\left( \begin{array}{c} \langle \dot \sigma_2 \rangle \\ \langle \dot \sigma_3 \rangle \end{array} \right)
&=& 2 \xi \left( \begin{array}{cc} 0 & -1  \\ 1 & 0 \end{array} \right)
\left( \begin{array}{c} \langle \sigma_2 \rangle \\ \langle \sigma_3 \rangle \end{array} \right) 
\end{eqnarray}
which can be solved easily analytically. 

A more straightforward way of solving the time evolution of the system is to write its state vector as $|\psi \rangle = \sum_{i=0,1} c_i \, |i \rangle$ and to use the Schr\"odinger equation to obtain differential equations for the complex coefficients $c_i$. However, the above approach of deriving rate equations for expectation values is more efficient, since we are only interested in the leakage of population out of the controlled subspace. Moreover, this approach can be extended easily to include more complex level schemes as well as the effect of spontaneous photon emission.

Since $\sigma_3$ commutes with $H_0$, we can calculate $P_0$ using the relation 
\begin{eqnarray} \label{P0unpro}
P_0 (t) = \frac{1}{2}(1+\langle \sigma_3 (t) \rangle) \, .
\end{eqnarray}
Solving Eq.~(\ref{rateq}) for time-independent coupling constants $\xi$ and for the case where the system is initially in $|0 \rangle$, we find that the population in the initial state changes according to
\begin{eqnarray}
P_0 (t) = \frac{1}{2} \left( 1 + \cos (2 \xi t) \right) = \cos^2 (\xi t) \, .  
\end{eqnarray}
This means, in the absence of any protection, the system remains inside the controlled subspace only on a time scale which is short compared to $1/\xi$. 

\section{Protecting a subspace with strong interactions} \label{pro1}

\begin{figure}[t]
\begin{center}
\resizebox{20pc}{!}{\rotatebox{-90}{\includegraphics{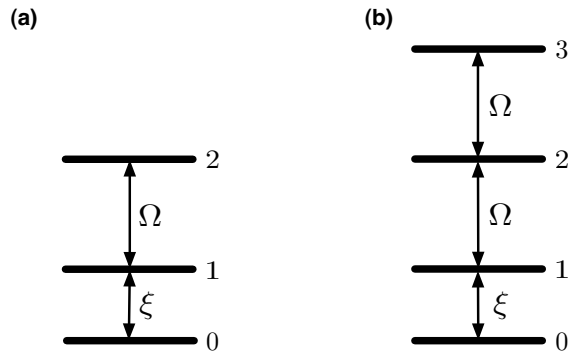}}}
\end{center}
\caption{Toy models to illustrate the possible protection of a controlled subspace (represented by $|0\rangle$) with strong interactions  with coupling strength $\Omega$ in the outside space. Here the outside space contains either the two states $|1 \rangle$ and $|2 \rangle$ (a) or the three states $|1 \rangle$, $|2 \rangle$, and $|3 \rangle$ (b).} \label{tony_2}
\end{figure}

One way to protect the controlled subspace against errors is to involve the relevant outside states into a relatively fast time evolution. Indeed it has been found that strong interactions can have the same effect as rapidly repeated measurements whether the system remains in its initial subspace or not \cite{Facchi4,Facchi3}. In good agreement with the predictions of the quantum Zeno effect \cite{misra}, these measurements strongly inhibit transitions out of the controlled subspace. In the following, we illustrate this approach with the help of the two toy models shown in Fig.~\ref{tony_2}. The purpose of the interactions with amplitude $\Omega$ is to induce fast oscillations of the amplitude of the state $|1 \rangle$. These cause $\langle \sigma_2 \rangle$ in Eq.~(\ref{defq}) to oscillate rapidly in time, such that $\langle \dot \sigma_3 \rangle$ in Eq.~(\ref{rateq}) becomes zero on average and the system remains approximately in $|0 \rangle$. 

As we shall see below, this strategy works well for the level scheme in Fig.~\ref{tony_2}(a). However, strong interactions acting on the outside space do not always protect the controlled subspace against leakage errors. Problems arise for example in the level scheme in Fig.~\ref{tony_2}(b). There the interactions in the outside space are more complex than the interactions which cause the leakage. The result is that the generation of approximate dark states in the outside space. These are zero eigenvectors of the fast system dynamics. Transitions between dark states and the controlled subspace are hence not protected by time scale separation, even when $\Omega$ becomes very large.

\subsection{Single-coupling case} \label{singcas}

We begin with an analysis of the three-level system shown in Fig.~\ref{tony_2}(a). Again, the controlled subspace contains only a single state, $|0 \rangle$, while the outside subspace contains the two states $|1 \rangle$ and $|2 \rangle$. In order to maximise the effect of the applied interactions, we assume resonant couplings. As before, $\xi$ is the coupling constant for the 0--1 transition, while $\Omega$ denotes the coupling constant for the 1--2 transition. Here we are especially interested in the case, where $\xi \ll \Omega$. Again we have a closer look at the time evolution of the population $P_0$ in the controlled subspace.

As in Section \ref{unpro}, we denote the energy of level $i$ by $\hbar \omega_i$. The Hamiltonian for the level configuration in Fig.~\ref{tony_2}(a) can then be written as
\begin{eqnarray}
H &=& \hbar \xi \, {\rm e}^{{\rm i}\omega_{\xi}t} \, |0 \rangle \langle 1| + \hbar \Omega \, {\rm e}^{{\rm i}\omega_{\Omega}t} \, |1 \rangle \langle 2| + {\rm h.c.} +  \sum_{i=0}^2 \hbar \omega_i \,  |i \rangle \langle i| 
\end{eqnarray}
with $\omega_\xi \equiv \omega_1 - \omega_0$ and $\omega_\Omega \equiv \omega_2 - \omega_1$. Transforming this Hamiltonian into the interaction picture with respect to $H_0 = \sum_{i=0}^2 \hbar \omega_i \,  |i \rangle \langle i|$, we obtain 
\begin{eqnarray} \label{HI31}
H_{\rm I} &=& \hbar \xi \, |0 \rangle \langle 1| + \hbar \Omega \, |1 \rangle \langle 2| + {\rm h.c.}
\end{eqnarray}
This interaction Hamiltonian is time independent and contains only the weak coupling between $|0 \rangle$ and $|1 \rangle$ and the strong coupling between $|1 \rangle$ and $|2 \rangle$. 

In order to obtain a closed system of rate equations, we now consider the expectation values of the Gell-Mann matrices \cite{Gell-Mann}
\begin{eqnarray} \label{Gell}
	&& \hspace*{-1cm} \sigma_2 = -{\rm i}\left( \ket{0}\bra{1} - \ket{1}\bra{0} \right) \, , ~~
	\sigma_7 = -{\rm i} \left( \ket{1}\bra{2} - \ket{2}\bra{1} \right) \, , ~~
	\sigma_4 = \ket{0}\bra{2} + \ket{2}\bra{0} \, , \nonumber \\
	&& \hspace*{-1cm}  \sigma_3 = \ket{0}\bra{0} - \ket{1}\bra{1} \, , ~~
	\sigma_8 = \frac{1}{\sqrt{3}}(\ket{0}\bra{0}+\ket{1}\bra{1}-2\, \ket{2}\bra{2}) \, .
\end{eqnarray}	
These are generalisations of the Pauli operators used in Section \ref{unpro}. Overall there are eight Gell-Mann matrices which can be used to model the time evolution of coupled three-level systems in a convenient way. However, due to the specific form of the interactions in the level scheme in Fig.~\ref{tony_2}(a), we need to consider only five of them. Using relation (\ref{dotA}), we find the following closed system of differential equations 
\begin{eqnarray} \label{rateq1}
\left( \begin{array}{c} \langle \dot \sigma_2 \rangle \\ \langle \dot \sigma_3 \rangle \\ \langle \dot \sigma_4 \rangle \\ \langle \dot \sigma_7 \rangle \\ \langle \dot \sigma_8 \rangle \end{array} \right)
&=& \left( \begin{array}{ccccc} 0 & -2 \xi & - \Omega & 0 & 0 \\ 2 \xi & 0 & 0 & - \Omega & 0 \\ \Omega & 0 & 0 & - \xi & 0 \\ 0 & \Omega & \xi & 0 & - \sqrt{3} \Omega \\ 0 & 0 & 0 & \sqrt{3} \Omega & 0 \end{array} \right)
\left( \begin{array}{c} \langle \sigma_2 \rangle \\ \langle \sigma_3 \rangle \\ \langle \sigma_4 \rangle \\ \langle \sigma_7 \rangle \\ \langle \sigma_8 \rangle \end{array} \right) \, .
\end{eqnarray}
These differential equations can be solved for example by calculating analytical expressions for the eigenvalues of this matrix or by simply using {\sl Mathematica}.

\begin{figure}[t]
\begin{center}
\resizebox{20pc}{!}{\rotatebox{-90}{\includegraphics{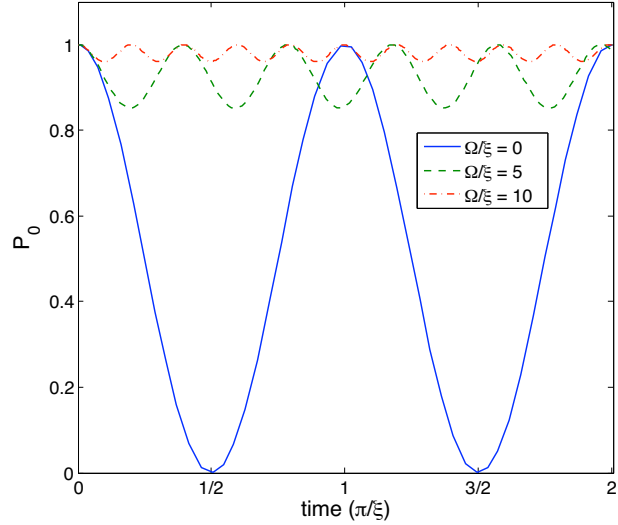}}}
\end{center}
\caption{Time dependence of $P_0$ for the level scheme shown in Fig.~\ref{tony_2}(a) for different ratios of $\Omega/\xi$. The system is initially in $|0 \rangle$. For $\Omega = 0$, the system leaves its initial state space on a time scale given by $1/\xi$. For $\Omega > 10 \, \xi$, the system remains there with a fidelity above $95 \, \%$ which constitutes an effective protection of the initial state space.} \label{P0_3}
\end{figure}

From Eq.~(\ref{Gell}) we see that the population in the controlled subspace equals
\begin{eqnarray} \label{P0three}
	 P_0 &=& \frac{1}{3} + \frac{1}{2} \, \langle \sigma_3 \rangle + \frac{1}{2\sqrt{3}} \, \langle \sigma_8 \rangle  \, .
\end{eqnarray}
Substituting the solution of the above rate equations into this equation, we find that the population in $|0 \rangle$ at time $t$ evolves according to
\begin{eqnarray} \label{P0_for_three}
	P_0 (t) &=& \frac{2\Omega^4 + \xi^4}{2\mu^4} + \frac{2 \Omega^2 \xi^2}{\mu^4}\cos (\mu t) + \frac{\xi^4}{2\mu^4}\cos (2\mu t)
\end{eqnarray}
with $\mu^2 \equiv \Omega^2 + \xi^2$, if the system was initially in $|0 \rangle$. For $\xi \ll \Omega$, Eq.~(\ref{P0_for_three}) simplifies to 
\begin{eqnarray} 
	P_0 (t) &=& 1 - \frac{2 \xi^2}{\Omega^2} \left[1 - \cos \left( \Omega t \right) \right]
\end{eqnarray}
which holds up to first order in $\xi^2/\Omega^2$. As shown in Fig.~\ref{P0_3}, the system remains to a very good approximation, i.e.~up to variations with an amplitude proportional to $\xi^2 / \Omega^2$, in $|0 \rangle$. This means, for $\xi^2 \ll \Omega^2$, the controlled subspace is effectively protected against leakage errors.

\subsection{Double-coupling case}

Using the same notation as in the previous subsection, the Hamiltonian for the level configuration in Fig.~\ref{tony_2}(b) in the Schr\"odinger picture equals   
\begin{eqnarray}
H &=&  \hbar \xi \, {\rm e}^{{\rm i}\omega_{\xi}t} |0 \rangle \langle 1| + \hbar \Omega \, {\rm e}^{{\rm i}\omega_{\Omega}t} \left( |1 \rangle \langle 2| + |2 \rangle \langle 3| \right) + {\rm h.c.} + \sum_{i=0}^3 \hbar \omega_i \,  |i \rangle \langle i| \, .
\end{eqnarray}
Again we first simplify this Hamiltonian by changing into the interaction picture with respect to the free Hamiltonian $H_0 = \sum_{i=0}^3 \hbar \omega_i \,  |i \rangle \langle i|$. This yields
\begin{eqnarray} \label{HI_4}
H_{\rm I} &=&  \hbar \xi \, |0 \rangle \langle 1| + \hbar \Omega \, \left( |1 \rangle \langle 2| + |2 \rangle \langle 3| \right) + {\rm h.c.} 
\end{eqnarray}
Instead of solving the corresponding Schr\"odinger equation, we apply again Eq.~(\ref{dotA}) to obtain a closed set of rate equations. 

To predict the time evolution of the population $P_0$ in the controlled subspace, we now have to consider nine generalised Gell-Mann matrices \cite{Gell-Mann}. These are  
\begin{eqnarray} \label{GM}
	&& \sigma_2 = -{\rm i} \, \left( \ket{0}\bra{1} - \ket{1}\bra{0} \right) \, , ~~ 
	\sigma_7 = -{\rm i} \, \left( \ket{1}\bra{2} - \ket{2}\bra{1} \right) \, , \nonumber \\
	&& \sigma_{10} = -{\rm i} \left( \ket{0}\bra{3} - \ket{3}\bra{0} \right) \, , ~~
	\sigma_{14} = -{\rm i} \, \left( \ket{2}\bra{3} - \ket{3}\bra{2} \right) \, , \nonumber \\
	&& \sigma_4 = \ket{0}\bra{2} + \ket{2}\bra{0} \, , ~~ 
	\sigma_{11} = \ket{1}\bra{3} + \ket{3}\bra{1} \, ,  \nonumber \\
	&& \sigma_3 = \ket{0}\bra{0} - \ket{1}\bra{1} \, , ~~
	\sigma_8 = \frac{1}{\sqrt{3}} \left( \ket{0}\bra{0} + \ket{1}\bra{1} - 2 \, \ket{2}\bra{2} \right) \, , ~~	   
	 \nonumber \\
	&& \sigma_{15} = \frac{1}{\sqrt{6}} \left( \ket{0}\bra{0} + \ket{1}\bra{1} + \ket{2}\bra{2} - 3\, \ket{3}\bra{3} \right) \, .
\end{eqnarray} 
Moreover, we notice that the interaction Hamiltonian (\ref{HI_4}) can be written as
\begin{eqnarray}
	H_{\rm I} = \hbar \xi \, \sigma_1 + \hbar \Omega \, \left(\sigma_6 + \sigma_{13} \right) 
\end{eqnarray}
with
\begin{eqnarray} \label{GM2}
	&& \sigma_1 = \ket{0}\bra{1} + \ket{1}\bra{0} \, ,~~
	 \sigma_6 = \ket{1}\bra{2} + \ket{2}\bra{1} \, , ~~ 
	\sigma_{13} = \ket{2}\bra{3} + \ket{3}\bra{2} \, .
\end{eqnarray} 
Substituting Eqs.~(\ref{GM})--(\ref{GM2}) into Eq.~(\ref{dotA}) and evaluating the relevant commutators, we see that the expectation of the operators in Eq.~(\ref{GM}) evolve according to 
\begin{eqnarray} \label{diffeq_4}
\hspace*{-3cm} \left( \begin{array}{c} \langle \dot \sigma_2 \rangle \\ \langle \dot \sigma_3 \rangle \\ \langle \dot \sigma_4 \rangle \\ \langle \dot \sigma_7 \rangle \\ \langle \dot \sigma_8 \rangle \\  \langle \dot \sigma_{10} \rangle \\ \langle \dot \sigma_{11} \rangle \\ \langle \dot \sigma_{14} \rangle \\ \langle \dot \sigma_{15} \rangle \end{array} \right)
&=& \left( \begin{array}{ccccccccc} 0 & -2 \xi & - \Omega & 0 & 0 & 0 & 0 & 0 & 0 \\ 
2 \xi & 0 & 0 & - \Omega & 0 & 0 & 0 & 0 & 0 \\ 
\Omega & 0 & 0 & - \xi & 0 & \Omega & 0 & 0 & 0 \\ 
0 & \Omega & \xi & 0 & - \sqrt{3} \Omega & 0 & - \Omega & 0 & 0 \\ 
0 & 0 & 0 & \sqrt{3} \Omega & 0 & 0 & 0 & - {2 \over \sqrt{3}} \Omega & 0 \\
0 & 0 & - \Omega & 0 & 0 & 0 & \xi & 0 & 0 \\
0 & 0 & 0 & \Omega & 0 & - \xi & 0 & - \Omega & 0 \\
0 & 0 & 0 & 0 & {2 \over \sqrt{3}} \Omega & 0 & \Omega & 0 & - {2 \sqrt{2} \over \sqrt{3}} \Omega \\
0 & 0 & 0 & 0 & 0 & 0 & 0 & {2 \sqrt{2} \over \sqrt{3}} \Omega & 0 \end{array} \right)
\left( \begin{array}{c} \langle \sigma_2 \rangle \\ \langle \sigma_3 \rangle \\ \langle \sigma_4 \rangle \\ \langle \sigma_7 \rangle \\ \langle \sigma_8 \rangle \\ \langle \sigma_{10} \rangle \\ \langle \sigma_{11} \rangle \\ \langle \sigma_{14} \rangle \\ \langle \sigma_{15} \rangle \end{array} \right) \, . \nonumber \\
\end{eqnarray}
This system of linear differential equations can, in principle, be solved analytically. However, for simplicity, we restrict ourselves to the presentation of a numerical solution. 

\begin{figure}[t]
\begin{center}
\resizebox{20pc}{!}{\rotatebox{-90}{\includegraphics{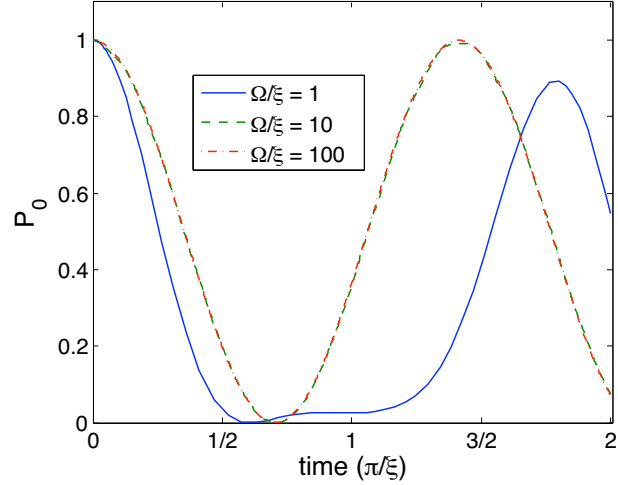}}}
\end{center}
\caption{Time dependence of $P_0$ for the level scheme shown in Fig.~\ref{tony_2}(b) for different ratios of $\Omega/\xi$. Here the controlled subspace is no longer protected against leakage, even when $\Omega$ becomes as large as $100 \, \xi$. The reason is the leakage of population into the dark state $|\lambda_0 \rangle$ which is illustrated in Fig.~\ref{4_level_sym}.} \label{P0_4}
\end{figure}

Using the Gell Mann matrices defined in Eq.~(\ref{GM}), the population in the initial state $P_0$ can now be written as 
\begin{eqnarray} \label{P0_for_4}
	 P_0 &=& \frac{1}{4} + \frac{1}{2} \, \langle \sigma_3 \rangle + \frac{1}{2\sqrt{3}} \, \langle \sigma_8 \rangle + \frac{1}{2\sqrt{6}} \, \langle \sigma_{15} \rangle \, .
\end{eqnarray}
The time evolution of $P_0$ obtained from substituting the numerical solution of the differential equation (\ref{diffeq_4}) into Eq.~(\ref{P0_for_4}) is shown in Fig.~\ref{P0_4}. Comparing the result for different values of $\Omega/\xi$ with the time evolution in the $\Omega=0$ case, we see that the controlled subspace is not protected, even when $\Omega$ is much larger than $\xi$. Leakage of population out of the controlled subspace happens on the same time scale as in the unprotected case.

\begin{figure}[t]
\begin{center}
\resizebox{15pc}{!}{\rotatebox{-90}{\includegraphics{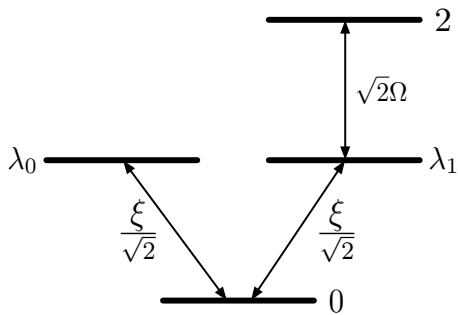}}}
\end{center}
\caption{Illustration of the effect of the Hamiltonian (\ref{HI_4_sym}) onto the states involved in the time evolution of the system. This level scheme is identical to the one shown in Fig.~\ref{tony_2}(b) but now we clearly see why the initial state $|0 \rangle$ is no longer protected against leakage errors.} \label{4_level_sym}
\end{figure}

Why does the protection of the controlled subspace work in the level scheme shown in Fig.~\ref{tony_2}(a) but not in the very similar level scheme shown in Fig.~\ref{tony_2}(b)? The reason for this becomes clear when we rewrite the Hamiltonian in Eq.~(\ref{HI_4}) in terms of the states $|0 \rangle$, 
\begin{eqnarray} \label{states}
	\ket{\lambda_0} \equiv \frac{1}{\sqrt{2}} (\ket{1} - \ket{3}) \, , ~~
	\ket{\lambda_1} \equiv \frac{1}{\sqrt{2}} (\ket{1} + \ket{3}) 
\end{eqnarray}
and $|2 \rangle$. Using this notation, $H_{\rm I}$ becomes
\begin{eqnarray} \label{HI_4_sym}
H_{\rm I} &=& {1 \over \sqrt{2}} \hbar \xi \, |0 \rangle \langle \lambda_0| 
+ {1 \over \sqrt{2}} \hbar \xi \, |0 \rangle \langle \lambda_1| 
+ \sqrt{2} \hbar \Omega \, |\lambda_1 \rangle \langle 2| + {\rm h.c.} 
\end{eqnarray}
The effect of this Hamiltonian is illustrated in Fig.~\ref{4_level_sym}. It shows that the system is only protected against leakage into the $|\lambda_1 \rangle$ state, since this state experiences a strong interaction. However, the system is not protected against leakage into $|\lambda_0 \rangle$, since $|\lambda_0 \rangle$ is a zero eigenstate of the $\Omega$ terms in Eq.~(\ref{HI_4_sym}). This means, $|\lambda_0 \rangle$ is not involved in a fast evolution and the transfer from $|0 \rangle$ to $|\lambda_0 \rangle$ occurs on the same time scale as in the unprotected case. In the final section, we show that dissipation is able to remove such dark states from the system so that the controlled subspace becomes protected again.

\section{Protecting a subspace with dissipation} \label{dissi}

\begin{figure}[t]
\begin{center}
\resizebox{20pc}{!}{\rotatebox{-90}{\includegraphics{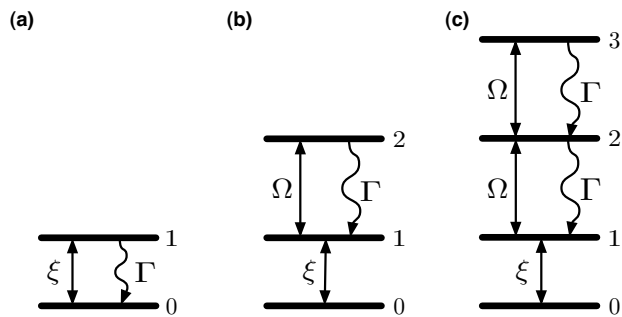}}}
\end{center}
\caption{Toy models to illustrate the possible protection of the controlled subspace (represented by $|0\rangle$) with a non-zero spontaneous decay rate $\Gamma$ and strong interactions with coupling strength $\Omega$ in the outside space.} \label{diss_level}
\end{figure}

In this section we analyse three examples (cf.~Fig.~\ref{diss_level}) which illustrate the possible protection of the controlled subspace {\em  using dissipation}. The controlled subspace contains again only the $|0 \rangle$ state, while the outside space contains one, two or three states. The only difference to the examples discussed in Sections \ref{unpro} and \ref{pro1} is the presence of a non-zero spontaneous decay rate $\Gamma$. As we shall see below, the controlled subspace is well protected against leakage in all three scenarios, when the interactions in the outside space described by $\Omega$ and the spontaneous decay rate $\Gamma$ are sufficiently larger than $\xi$. 

\subsection{Single-coupling case with dissipation} \label{sincas2}

Let us first have a look at the three-level system shown in Fig.~\ref{diss_level}(b). To describe its time evolution, we go again into the interaction picture with respect to the free evolution and analyse the master equation
\begin{eqnarray} \label{master_2}
	\dot{\rho} = - {{\rm i} \over \hbar} [ \, H_{\rm I} , \rho \, ] + \frac{\Gamma}{2} \Big[ \, 2 \, \ket{1}\bra{2} \, \rho \, \ket{2}\bra{1} - \rho \, \ket{2}\bra{2} - \ket{2}\bra{2} \, \rho \, \Big] \, .
\end{eqnarray}
The interaction Hamiltonian $H_{\rm I}$ is the same as in Eq.~(\ref{HI31}). In order to predict the time evolution of the population in the controlled subspace, we derive again a closed system of rate equations. The time derivative of the expectation value of an operator $A$ is now given by 
\begin{eqnarray} \label{dot_sigma}
	\expect{\dot{A}} = {\rm Tr} \left(A \dot{\rho}\right) \, .
\end{eqnarray}
Taking this into account, we find that the Gell Mann matrices in Eq.~(\ref{Gell}) evolve according to 
\begin{eqnarray} \label{rateq11}
\hspace*{-1.5cm} \left( \begin{array}{c} \langle \dot \sigma_2 \rangle \\ \langle \dot \sigma_3 \rangle \\ \langle \dot \sigma_4 \rangle \\ \langle \dot \sigma_7 \rangle \\ \langle \dot \sigma_8 \rangle \end{array} \right)
&=& \left( \begin{array}{ccccc} 0 & -2 \xi & - \Omega & 0 & 0 \\ 
2 \xi & 0 & 0 & - \Omega & {1 \over \sqrt{3}} \Gamma \\ 
\Omega & 0 & - {1 \over 2} \Gamma & - \xi & 0 \\ 
0 & \Omega & \xi & - {1 \over 2} \Gamma & - \sqrt{3} \Omega \\ 
0 & 0 & 0 & \sqrt{3} \Omega & - 3 \Gamma \end{array} \right)
\left( \begin{array}{c} \langle \sigma_2 \rangle \\ \langle \sigma_3 \rangle \\ \langle \sigma_4 \rangle \\ \langle \sigma_7 \rangle \\ \langle \sigma_8 \rangle \end{array} \right) + \left( \begin{array}{c} 0 \\ - {1 \over 3} \Gamma \\ 0 \\ 0 \\ \sqrt{3} \Gamma \end{array} \right)
\, .
\end{eqnarray}
These equations resemble the ones shown in Eq.~(\ref{rateq1}). The additional $\Gamma$ terms take the effect of dissipation into account. 

\begin{figure}[t]
\begin{center}
\resizebox{20pc}{!}{\rotatebox{-90}{\includegraphics{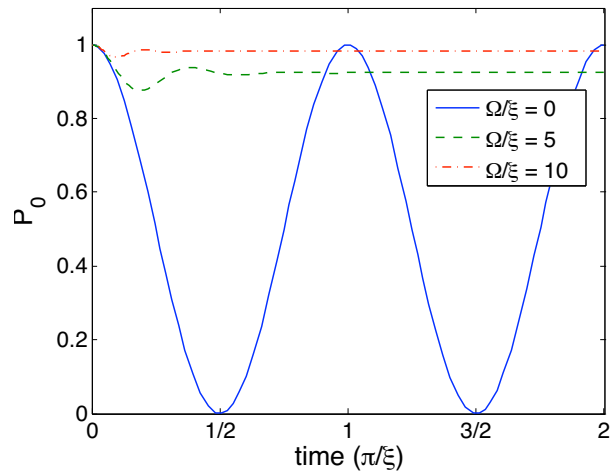}}}
\end{center}
\caption{Time dependence of $P_0$ for the level scheme shown in Fig.~\ref{diss_level}(b) for $\Gamma = \Omega$ and different ratios of $\Omega/\xi$. On average, the protection of the controlled subspace is more or less the same as in Fig.~\ref{P0_3} which corresponds to the same level scheme but with $\Gamma=0$ (cf.~Fig.~\ref{tony_2}(a)).} \label{P0_3_diss}
\end{figure}

The population in the controlled subspace can be obtained by substituting for example the numerical solution of these equations into Eq.~(\ref{P0three}). The result is shown in Fig.~\ref{P0_3_diss}. For simplicity we assumed $\Gamma = \Omega$. For $\Omega=0$, we see again the Rabi oscillations of the unprotected case. However, when $\Omega$ becomes sufficiently larger than $\xi$, then the system remains to a very good approximation in its initial state. On average, the protection of the controlled subspace is more or less the same as in Section \ref{singcas}, where we had $\Gamma=0$ (cf.~Fig.~\ref{P0_3}). There seems to be no advantage of having a non-zero spontaneous decay rate in the system! Notice that having $\Gamma \neq 0$ in the level scheme in Fig.~\ref{diss_level}(b) is only advantageous when someone actually observes whether the system emits photons or not. Indeed, one can show that the system remains in its initial state $|0 \rangle$ with a very high fidelity under the condition of {\em no} photon emission \cite{BeHe}. If a photon emission is detected, then the system has left the controlled subspace and the anticipated control experiment needs to be restarted. 

Comparing the level scheme in Fig.~\ref{diss_level}(b) with the level scheme analysed in Refs.~\cite{BeHe,Dehmelt,MQJ}, we see that its dynamics exhibits so-called macroscopic light and dark periods. Indeed, for $\xi$ much smaller than $\Omega$ and $\Gamma$, the initial state $|0 \rangle$ is an approximate zero eigenstate of the system dynamics. The absence of photon emissions hence confirms that the system is in this state. As a consequence of the quantum Zeno effect, it therefore remains there for a relatively long time. On average, this time equals $\Omega^2/\Gamma \xi^2$ which is much larger than $1/\xi$ \cite{BeHe}. In other words, the system exhibits a macroscopic dark period. The system may eventually drop out of the controlled subspace, thereby entering a so-called macroscopic light period and causing fluorescence at a rate which depends on $\Omega$ and $\Gamma$. This behaviour is not reflected in Fig.~\ref{P0_3_diss}, since the density matrix description used in this paper does not allow us to distinguish the different trajectories of the system.

\subsection{Single-state outside} \label{2_level_2}

\begin{figure}[t]
\begin{center}
\resizebox{19.5pc}{!}{\rotatebox{-90}{\includegraphics{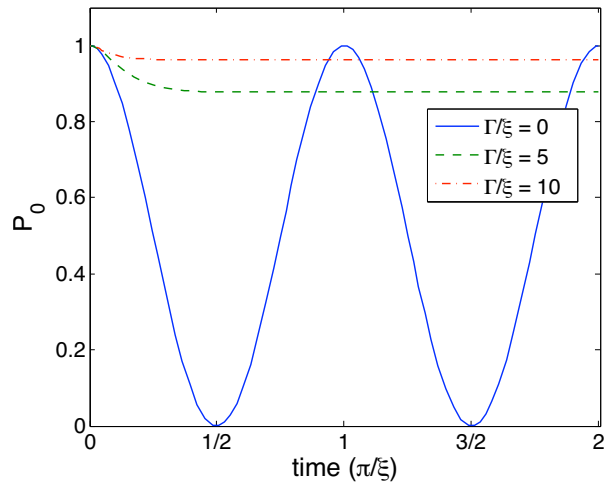}}}
\end{center}
\caption{Time dependence of $P_0$ for the level scheme shown in Fig.~\ref{diss_level}(a) for different ratios of $\Gamma/\xi$. For $\Gamma \gg \xi$, the system remains in the controlled subspace with a very high fidelity.} \label{P0_2_diss}
\end{figure}

Let us now have a look at the simple level configuration in Fig.~\ref{diss_level}(a). Its time evolution is given by the master equation 
\begin{eqnarray} \label{master}
	\dot{\rho} = - {{\rm i} \over \hbar} [ \, H_{\rm I} , \rho \, ] + \frac{\Gamma}{2} \Big[ \, 2 \, \ket{0}\bra{1} \, \rho \, \ket{1}\bra{0} - \rho \, \ket{1}\bra{1} - \ket{1}\bra{1} \, \rho \, \Big] \, .
\end{eqnarray}
In the interaction picture with respect to the free evolution, the interaction Hamiltonian $H_{\rm I}$ is the same as in Eq.~(\ref{HI_1}). To predict the time evolution of $P_0$ we proceed as in Section \ref{unpro} and consider again the Pauli operators in Eq.~(\ref{defq}). Their expectation values evolve now according to
\begin{eqnarray}
	\left( \begin{array}{c} \expect{ \dot \sigma_2 } \\ \expect{ \dot \sigma_3 } \end{array} \right)
	&=& \left( \begin{array}{cc} -\frac{1}{2}\Gamma & -2 \xi \\ 
	2 \xi & -\Gamma \end{array} \right)
	\left( \begin{array}{c} \expect{\sigma_2} \\ \expect{\sigma_3} \end{array} \right) + \left( \begin{array}{c} 0 \\ \Gamma \end{array} \right) \, .
\end{eqnarray}
Combining the result of this equation with Eq.~(\ref{P0unpro}) yields the time dependence of the population $P_0$ in the controlled subspace. 

Fig.~\ref{P0_2_diss} shows a numerical solution of the time dependence of $P_0$ for different ratios $\Gamma/\xi$. For $\Gamma=0$, we observe the Rabi oscillations in and out of the initial subspace which occur in the unprotected case. For $\Gamma \gg \xi$, the state vector becomes $|0 \rangle$ with a very high fidelity. But even for relatively modest values for $\Gamma/\xi$, the density matrix $\rho$ settles quickly into a steady state with the system predominantly in $|0 \rangle$. The reason for this very strong protection of the controlled subspace is that, even when it leaves, the system returns very rapidly via the spontaneous emission of a photon. 

\subsection{Double-coupling case with dissipation}

\begin{figure}[t]
\begin{center}
\resizebox{20pc}{!}{\rotatebox{-90}{\includegraphics{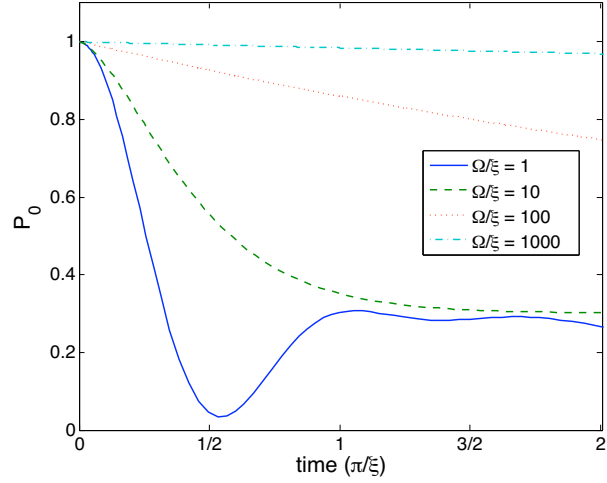}}}
\end{center}
\caption{Time dependence of $P_0$ for the level scheme shown in Fig.~\ref{diss_level}(c) for $\Gamma = \Omega$ and different ratios of $\Omega/\xi$. Compared to Fig.~\ref{P0_4}, we now observe an increasing effectiveness of protection of the controlled subspace with increasing values of $\Omega$.} \label{P0_4_diss}
\end{figure}

This final subsection analyses the time evolution of the four-level system shown in Fig.~\ref{diss_level}(c). The only difference to the level scheme in Fig.~\ref{tony_2}(b) is the presence of the non-zero spontaneous decay rate $\Gamma$. To calculate the time evolution of the population in the controlled subspace, we now consider the master equation
\begin{eqnarray} \label{master_3}
	\dot{\rho} &=& - {{\rm i} \over \hbar} [ \, H_{\rm I} , \rho \, ] + \frac{\Gamma}{2} \Big[ \, 2 \, \ket{1}\bra{2} \, \rho \, \ket{2}\bra{1} - \rho \, \ket{2}\bra{2} - \ket{2}\bra{2} \, \rho \, \Big] \nonumber \\
	&& + \frac{\Gamma}{2} \Big[ \, 2 \, \ket{2}\bra{3} \, \rho \, \ket{3}\bra{2} - \rho \, \ket{3}\bra{3} - \ket{3}\bra{3} \, \rho \, \Big] \, ,
\end{eqnarray}
whose interaction Hamiltonian $H_{\rm I}$ can be found in Eq.~(\ref{HI_4}). Proceeding as above, we find that the time evolution of the Gell Mann matrices (\ref{GM}) is now given by the differential equations
\begin{eqnarray} \label{diffeq_4_diss}
&& \hspace*{-1.5cm} \left( \langle \dot \sigma_2 \rangle, \, \langle \dot \sigma_3 \rangle, \,  \langle \dot \sigma_4 \rangle, \,  \langle \dot \sigma_7 \rangle, \,  \langle \dot \sigma_8 \rangle, \,   \langle \dot \sigma_{10} \rangle, \, \langle \dot \sigma_{11} \rangle, \,  \langle \dot \sigma_{14} \rangle, \,  \langle \dot \sigma_{15} \rangle \right)^{\rm T} \nonumber \\
&=& M \left( \langle \sigma_2 \rangle , \,  \langle \sigma_3 \rangle, \,  \langle \sigma_4 \rangle, \,  \langle \sigma_7 \rangle, \, \langle \sigma_8 \rangle, \,  \langle \sigma_{10} \rangle, \,  \langle \sigma_{11} \rangle, \,  \langle \sigma_{14} \rangle , \, \langle \sigma_{15} \rangle \right)^{\rm T} \nonumber \\
&& + \left( 0 , \,  - {1 \over 4} \Gamma , \,  0 , \,  0, \, {1 \over 4 \sqrt{3}} \Gamma , \,  0 , \, 0 , \,  0 , \, {1 \over \sqrt{6}} \Gamma \right)^{\rm T} 
\end{eqnarray}
with
\begin{eqnarray} \label{diff_mat}
\hspace*{-1.5cm} M &=& \left( \begin{array}{ccccccccc} 
0 & -2 \xi & - \Omega & 0 & 0 & 0 & 0 & 0 & 0 \\ 
2 \xi & 0 & 0 & - \Omega & {1 \over \sqrt{3}} \Gamma & 0 & 0 & 0 & - {1 \over 2 \sqrt{6}} \Gamma \\ 
\Omega & 0 & - {1 \over 2} \Gamma & - \xi & 0 & \Omega & 0 & 0 & 0 \\ 
0 & \Omega & \xi & - {1 \over 2} \Gamma  & - \sqrt{3} \Omega & 0 & - \Omega & 0 & 0 \\ 
0 & 0 & 0 & \sqrt{3} \Omega & - \Gamma & 0 & 0 & - {2 \over \sqrt{3}} \Omega & {3 \over 2 \sqrt{2}} \Gamma \\
0 & 0 & - \Omega & 0 & 0 & - {1 \over 2} \Gamma & \xi & 0 & 0 \\
0 & 0 & 0 & \Omega & 0 & - \xi & - {1 \over 2} \Gamma & - \Omega & 0 \\
0 & 0 & 0 & 0 & {2 \over \sqrt{3}} \Omega & 0 & \Omega & - \Gamma & - {2 \sqrt{2} \over \sqrt{3}} \Omega \\
0 & 0 & 0 & 0 & 0 & 0 & 0 & {2 \sqrt{2} \over \sqrt{3}} \Omega & - \Gamma \end{array} \right) \, .
\end{eqnarray}
Fig.~\ref{P0_4_diss} shows the time dependence of $P_0$ for the case where the system is initially in the controlled subspace and has been obtained by substituting the numerical solution of these equations into Eq.~(\ref{P0_for_4}).

Comparing Figs.~\ref{P0_4} and \ref{P0_4_diss}, we see that the presence of a sufficiently large spontaneous decay rate $\Gamma$ combined with the presence of a relatively large coupling constant $\Omega$ now results in an effective protection of the controlled subspace against leakage errors. There are different ways of seeing how this protection (which was not there before) has been achieved. One way is to have a closer look at the 
above master equation and to notice that the state $|\lambda_0 \rangle$ is no longer a zero eigenstate of the system dynamics. Whenever, population accumulates in this state, the system returns (either via the emission of a photon or as a result of its no-photon evolution) on the time scale given by $\Gamma$ into $|1 \rangle$, where it experiences fast driving with $\Omega$. This example confirms that dissipation can provide a very efficient tool for restricting the time evolution of a system onto a controlled subspace. 

Another way to gain an intuition into the behaviour of the level scheme in Fig.~\ref{diss_level}(c) is to compare it to the level scheme in Fig.~\ref{diss_level}(b) which we analysed in Section \ref{sincas2}. Observing whether the system emits photons or not, one would notice again two very distinct phases of operation. The system either emits photons at a high rate or it remains dark for a relatively long time. A macroscopic light period, on one hand, indicates that the state vector lies entirely outside the controlled subspace. A macroscopic dark period, on the other hand, indicates that the system is in $|0 \rangle$. In other words, if the system is initially in the controlled subspace, it remains there on average much longer than in the unprotected case. The result is the protection of the system against leakage errors which, when they occur, are heralded by an easy-to-detect fluorescence signal.
 
\section{Conclusions} \label{conc}

This paper illustrates two methods to protect a controlled subspace against the leakage of population into the outside space: one using strong interactions in the outside space and one using dissipation. This is done with the help of relatively simple toy models whose time evolution can be analysed relatively easily. For simplicity, we assume that the controlled subspace consists only of one state, namely $|0 \rangle$. The outside space contains either one, two or three states denoted $|1 \rangle$, $|2 \rangle$, and $|3 \rangle$ (cf.~Figs.~\ref{tony_2} and \ref{diss_level}). Section \ref{unpro} discusses the unprotected case and shows that unwanted transitions from $|0 \rangle$ to $|1 \rangle$ (due to resonant coupling) occur on a time scale given by a relatively small parameter $\xi$ (cf.~Fig.~\ref{tony_1}). 

In Section \ref{pro1}, the decoherence time of the system  is increased to one which scales as $\xi^2$ by applying relatively fast interactions with coupling strength $\Omega$ to the outside space. However, these strong interactions are not always sufficient for protecting a controlled subspace against leakage errors. While it works well for the level scheme shown in Fig.~\ref{tony_2}(a), no protection occurs for the level scheme shown in Fig.~\ref{tony_2}(b). The reason is the existence of an approximate zero eigenstate outside the controlled subspace. This state does not experience fast driving and therefore behaves as the state $|1 \rangle$ in the unprotected case.  

Section \ref{dissi} considers three scenarios where a spontaneous decay rate $\Gamma$ has been added to the 
level schemes analysed in Sections \ref{unpro} and \ref{pro1}. All the level schemes shown in Fig.~\ref{diss_level} exhibit a strong protection of the controlled subspace. One way to understand the mechanism which inhibits the population transfer out of the controlled subspace is to interpret the behaviour of the system in terms of the quantum Zeno effect \cite{misra,tony,zeno}. Suppose being outside the controlled subspace results necessarily in the spontaneous emission of a photon. Then, observing whether a photon emission takes place or not is equivalent to performing a measurement on whether the system is in the controlled subspace or not. If these measurements occur on a sufficiently short time scale, then a system initially in the controlled subspace remains there much longer than in an unobserved case. A similar interpretation applies to the protection of the controlled subspace with a strong interaction illustrated in Fig.~\ref{tony_2}(a) \cite{Facchi4,Facchi3}.

Finally, let us remark that the absence of decoherence within the controlled subspace results in general in an effective time evolution which can be described by the effective Hamiltonian \cite{Shore,viola,Beige00,Facchi5}
\begin{eqnarray}
H_{\rm eff} &=& {\bf P}_{\rm CS} \, H \, {\bf P}_{\rm CS} \, ,
\end{eqnarray}
where ${\bf P}_{\rm CS}$ denotes the projector onto the controlled subspace and $H$ is the total system Hamiltonian. Once the protection is in place, the interactions described by $H_{\rm eff}$ can be designed as required by the control task at hand. Applications of such control tasks can be found for example in quantum information processing, i.e.~in the realisation of gate operations and the preparation of highly entangled states. Recent ideas for achieving these tasks use dissipation in an even more constructive way, for example, by heralding successful state preparations with macroscopic fluorescence signals \cite{Metz,Metz2,Jonathan,meschede} or by letting non-unitary evolutions guide the system into the desired target states \cite{tannor,vacanti,schirmer2,Verstraete,Kraus,Viola}.

\ack
A. B. thanks Christiane Koch, Tony Sudbury, and Lorenza Viola for very interesting and valuable discussions. She also acknowledges a James Ellis University Research Fellowship from the Royal Society and the GCHQ. This work was moreover supported by the UK Research Council EPSRC through the QIP IRC and by the EU Research and Training Network EMALI.

\end{document}